\documentclass[aps,preprint,prl,superscriptaddress,amsmath,amssymb,amsfonts]{revtex4-2}
\usepackage{graphicx}  
\usepackage{color}
\usepackage[colorlinks,bookmarks=true,citecolor=blue,linkcolor=blue,urlcolor=blue, breaklinks=true]{hyperref}
\usepackage{bm}
\usepackage{amsmath,amssymb}
\usepackage{natbib}

\begin{document}

\title{Gapless spin excitations in a quantum spin liquid state of {\boldmath $S$}=1/2 perfect kagome antiferromagnet}

\author{S. Suetsugu}
\thanks{Corresponding authors.}
\affiliation{Department of Physics, Kyoto University, Kyoto 606-8502, Japan}
\author{T. Asaba}
\affiliation{Department of Physics, Kyoto University, Kyoto 606-8502, Japan}
\author{S. Ikemori}
\affiliation{Department of Physics, Kyoto University, Kyoto 606-8502, Japan}
\author{Y. Sekino}
\affiliation{Department of Physics, Kyoto University, Kyoto 606-8502, Japan}
\author{Y. Kasahara}
\affiliation{Department of Physics, Kyoto University, Kyoto 606-8502, Japan}
\author{K. Totsuka}
\affiliation{Yukawa Institute for Theoretical Physics, Kyoto University, Kyoto 606-8502, Japan}
\author{B. Li}
\affiliation{Wuhan National High Magnetic Field Center and School of Physics, Huazhong University of Science and Technology, 430074 Wuhan, China}
\author{Y. Zhao}
\affiliation{Wuhan National High Magnetic Field Center and School of Physics, Huazhong University of Science and Technology, 430074 Wuhan, China}
\author{Y. Li}
\affiliation{Wuhan National High Magnetic Field Center and School of Physics, Huazhong University of Science and Technology, 430074 Wuhan, China}
\author{Y. Kohama}
\affiliation{Institute for Solid State Physics, The University of Tokyo, Kashiwa, Chiba 277-8581, Japan}
\author{Y. Matsuda}
\thanks{Corresponding authors.}
\affiliation{Department of Physics, Kyoto University, Kyoto 606-8502, Japan}

\date{\today}



\begin{abstract}
Quantum spin liquids (QSLs) represent an exotic quantum many-body state characterized by the suppression of long-range magnetic order due to strong quantum fluctuations. The kagome spin-1/2 antiferromagnet (AFM) is a prime candidate for realizing QSLs, but its ground state remains an unresolved conundrum. Here we investigate the recently discovered perfect kagome AFM YCu$_3$(OH)$_{6.5}$Br$_{2.5}$ to elucidate two central enigmas surrounding the kagome AFM. Ultra-sensitive torque magnetometry experiments reveal that the intrinsic magnetic susceptibility arising from the kagome layer remains nearly temperature-independent down to exceedingly low temperatures. This observation seemingly implies the emergence of gapless fermionic spin excitations akin to Pauli paramagnetism in metals. However, most strikingly, these results stand in stark contrast to the conspicuous absence of a temperature-linear contribution to the specific heat. These findings appear irreconcilable with the widely-discussed theoretical frameworks assuming fermionic quasiparticles (QPs), instead suggesting a transition of bosonic QPs into a superfluid state with a gapless Goldstone mode. Furthermore, magnetocaloric measurements evince an entropy anomaly, constituting thermodynamic evidence that magnetic fields instigate the opening of a spin gap, driving a quantum phase transition into a 1/9 magnetization plateau state. These results shed light on the nature of the low-energy excitations in zero and strong magnetic fields, providing crucial insights into the long-standing unresolved issues of the ground state of the kagome AFM.
\end{abstract}

\maketitle

The antiferromagnetic spin-1/2 Heisenberg model on a two-dimensional (2D) kagome lattice with a corner-sharing triangular structure represents one of the simplest models, exhibiting strong frustration and quantum fluctuations. The synergistic interplay of geometrical frustration, antiferromagnetic interactions, and the quantum nature of $S$ = 1/2 spins is believed to stabilize a highly entangled QSL state \cite{PhysRevB.45.12377,balents2010spin}, engendering exotic properties and fractionalized excitations. However, despite longstanding intensive study, elucidating the ground state of the quantum kagome AFM has been one of the most vexing issues in quantum magnetism \cite{PhysRevLett.98.117205,yan2011spin,jiang2012identifying,PhysRevLett.109.067201,PhysRevB.87.060405,PhysRevX.7.031020,PhysRevB.100.155142}. The primary reason for the difficulty in identifying the ground state is the existence of a plethora of competing states that are exceedingly close in energy, leading to a wide range of theoretical proposals for the potential ground state. Indeed, the QSL states that may arise are manifold, encompassing the gapped $Z_2$ spin liquid \cite{yan2011spin,jiang2012identifying,PhysRevLett.109.067201,PhysRevB.100.155142} and the gapless $U(1)$ Dirac spin liquid \cite{PhysRevLett.98.117205,PhysRevB.87.060405,PhysRevX.7.031020}, and the specific QSL state manifested in the kagome AFM continues to elude precise identification.

Another outstanding open question in the kagome AFM pertains to the nature of the ground state when the system is subjected to extremely high magnetic fields. In particular, the magnetization curves reflect highly non-trivial magnetic states and excitations \cite{nishimoto2013controlling,PhysRevB.93.060407,chen2018thermodynamics,PhysRevB.107.L220401}. Of particular interest is whether the 1/9 magnetization plateau which originates from purely quantum mechanical effects, manifests in the kagome AFM. The unambiguous experimental observation of this elusive 1/9 plateau with a distinct spin gap would provide crucial insights into the quantum nature of the kagome AFM.

The realization and exquisite characterization of the kagome AFM have been pivotal endeavors in the quest to unveil the enigmatic QSL ground state. Herbertsmithite [ZnCu$_3$(OH)$_6$Cl$_2$] \cite{mendels2010quantum} has emerged as the first kagome AFM featuring a perfect kagome lattice of Cu$^{2+}$ ions ($S=1/2$) and dominant nearest-neighbor Heisenberg antiferromagnetic interactions $J$($\sim$180-190\,K). In this material, no magnetic order has been detected down to 50\,mK \cite{PhysRevLett.98.077204,PhysRevLett.98.107204}, and a spinon continuum has been observed through inelastic neutron scattering (INS) experiments \cite{han2012fractionalized}. However, the presence of orphan spins, introduced by the intersite mixing of Cu$^{2+}$ and Zn$^{2+}$ within and between the kagome planes \cite{shores2005structurally,freedman2010site}, significantly perturbed the low-energy excitations observed in the specific heat \cite{kimchi2018scaling,PhysRevB.106.174406} and magnetic susceptibility \cite{PhysRevB.76.132411,PhysRevLett.104.147201}. Nuclear magnetic resonance studies aimed at isolating impurity contributions have yielded controversial results regarding the presence/absence of a spin gap \cite{fu2015evidence,khuntia2020gapless,wang2021emergence}. Furthermore, the anticipated magnetic plateau behavior characteristic of kagome AFM has not been discernibly observed under extremely high magnetic fields \cite{han2014thermodynamic}.

The recently discovered spin-1/2 kagome AFM YCu$_3$(OH)$_{6.5}$Br$_{2.5}$ \cite{chen2020quantum,PhysRevB.105.024418,PhysRevB.105.L121109,lu2022observation} (denoted YCOB) presents a promising platform to investigate the intrinsic ground state of the kagome QSL. This compound features a perfect 2D kagome lattice of Cu$^{2+}$ ions [Figs.\,1(a) and (b)] and exhibits no magnetic order down to 50\,mK \cite{PhysRevB.105.L121109}, despite a sizable $J$ of $\sim$80\,K \cite{suetsugu2023emergent}. The absence of intersite mixing between magnetic Cu$^{2+}$ and non-magnetic Y$^{3+}$ ions effectively suppresses disorder effects arising from orphan spins and impurities. This characteristic enables the exploration of the intrinsic ground state and its associated excitations of the 2D kagome QSL, largely unobscured by disorder effects. Although the spin excitations have been investigated by specific heat \cite{PhysRevB.105.024418,PhysRevB.105.L121109} and INS \cite{zeng2023dirac} measurements, the presence/absence of the spin gap remains open. Moreover, INS measurements have evinced a linear dispersion relation for the QP excitations, $E \propto |\bm{k}|$, which has been deliberated in the context of Dirac fermions. However, further experiments probing lower energy scales are imperative to elucidate the spin gap conundrum and unveil the intrinsic nature of the QPs. When subjected to strong magnetic fields, YCOB exhibits plateau-like magnetization anomalies at 1/9 and 1/3 of the fully polarized value \cite{suetsugu2023emergent,jeon2024,zheng2023unconventional}. It should be stressed that this magnetization behavior implies that the magnetic interactions within the material are well-defined and not seriously affected by random perturbations or impurities. While the 1/3 plateau is widely accepted as a genuine plateau representing a quantum phase, the origin of the 1/9 anomaly remains contentious. In this study, we elucidate the highly nontrivial nature of the 2D kagome AFM in zero and intense magnetic fields by examining YCOB.

The temperature ($T$) dependence of the magnetic susceptibility $\chi(T) = M/H$, where $M$ is magnetization and $H$ is a magnetic field, is depicted in Fig.\,1(c). Below approximately 30\,K, $\chi(T)$ exhibits a notable increase. As will be discussed later, this enhancement can be attributed, in part, to the presence of impurity spins that show a Curie-like divergence, where $\chi(T)$ is proportional to $1/T$. The contribution from these impurity spins obfuscates the intrinsic $T$-dependence of the low-temperature susceptibility arising solely from the 2D kagome lattice. Consequently, it becomes exceedingly challenging to discern the potential presence of a spin gap from magnetic susceptibility measurements, even in YCOB with significantly reduced concentrations of impurity spins.

To resolve the intrinsic magnetic susceptibility $\chi_{K}$ solely from the QSL state in the kagome lattice, we performed magnetic torque $\bm{\tau}$ measurements using a highly sensitive capacitive cantilever \cite{li2011coexistence,PhysRevB.90.064417} by applying fields $\bm{H}$ at an angle $\theta$ relative to the $c$-axis [inset of Fig.\,2(a)]. The torque $\bm{\tau}$ is a thermodynamic quantity equal to the derivative of the free energy $F$ with respect to $\theta$, given by $\bm{\tau}= \partial F/\partial \theta = \mu_0V \bm{M} \times \bm{H}$, where $\mu_0$ is vacuum permeability, and $V$ is the sample volume. The amplitude of the torque is given by $\tau = \mu_0V(M_{ab}H_c - M_cH_{ab})$, where $M_{ab}$ and $M_{c}$ are the magnetization within the $ab$-plane and parallel to the $c$-axis, and $H_{ab}$ and $H_{c}$ are the corresponding field components, respectively. In the paramagnetic state, the $ab$-plane susceptibility $\chi_{ab}$ and the $c$-axis susceptibility $\chi_c$ are proportional to the magnetization components as $\chi_{ab} = M_{ab} / H_{ab}$ and $\chi_c = M_c / H_c$, respectively. Consequently, the amplitude of the torque is expressed as $\tau=\frac{1}{2}\mu_0VH^2\Delta\chi\sin 2\theta$, where $\Delta\chi = \chi_{ab} - \chi_c$ is the anisotropy of $\chi$. A significant advantage of this method is the suppression of substantial Curie-like contributions arising from impurity spins at low temperatures, provided that these impurities yield isotropic magnetic responses.

Figure\,2(a) depicts $\tau/H$ of sample \#1 at $\theta = 43^\circ$ as a function of $\mu_0H$ for fields below 1\,T and temperatures below 20\,K. In this low-field regime, $\tau/H$ exhibits a nearly linear increase with $H$ across the entire $T$-range. This linear $H$-dependence was corroborated in the same crystal at a different tilt angle ($\theta = 7^\circ$) and in a different crystal (sample \#2), as shown in Fig.\,2(b) (see also Fig.\,S1). These observations provide evidence both $M_{ab}$ and $M_c$ increase linearly with $H$, indicating the system is in the paramagnetic state in this low-field regime, where $\tau/H \propto H\Delta\chi$. The field slope of $\tau/H$ is almost independent of temperature below 6\,K, in contrast to the magnetization measured by SQUID magnetometry [Fig.\,1(d)]. These results demonstrate the Curie contribution arising from impurity spins is nearly perfectly canceled out in the torque measurements. Consequently, $\Delta\chi$ originates purely from the 2D kagome plane, and hereafter we denote $\Delta\chi_K \equiv \Delta\chi$. At higher fields, $\tau/H$ continues to increase up to 13\,T [Fig.\,S1(c)]. The slightly non-linear behavior observed at low temperatures likely reflects the detailed band dispersion of the spin excitations. It is noteworthy that in the case of Herbertsmithite with a large amount of orphan spins, the magnetic torque shows a pronounced non-linear behavior with respect to $H$ and a complex $T$-dependence, including a sign change \cite{PhysRevB.90.064417}.

Figure\,2(c) illustrates the $T$-dependence of $\Delta\chi_K (\propto \tau/H^2)$ at 0.5\,T. As temperature decreases, $\Delta\chi_K$ increases up to 5\,K and then saturates in the low-temperature regime. The $T$-independent behavior of $\Delta\chi_K$ below 5\,K indicates that $\Delta\chi_K$ remains finite down to the lowest measured temperature. Notably, this $T$-independent $\Delta \chi_K$ directly implies that the magnetic susceptibility intrinsic to the 2D kagome layers $\chi_K$ exhibits a $T$-independent behavior for both parallel and perpendicular to $H$. To quantitatively extract this finite value of $\chi_K$, we compare $\Delta\chi_K$ with $\chi$ measured by SQUID magnetometry [solid line in Fig.\,2(c)]. In the low-temperature limit, $\chi_K$ is estimated to be $6.5\times 10^{-3}$\,emu/mol-Cu.

We now discuss the issue of the spin gap more quantitatively. The results presented in Fig.\,2(b) demonstrate that both $M_{ab}$ and $M_c$ are proportional to $H$ at least down to sufficiently low magnetic fields of about 50\,mT, which corresponds to a temperature scale of 30\,mK. This is well below the lowest measured temperature (160\,mK). Therefore, we can conclude that the spin gap, if it exists, is sufficiently below 160\,mK, i.e., less than 1/500 of the energy scale of the exchange interaction ($J \sim$80\,K). Thus, the present results provide thermodynamic evidence that the spin gap is either absent or undetectably small in the zero-field limit in YCOB.

Specific heat $C/T$ measurements (Fig.\,3) provide further insights into the gapless excitations. The zero-field specific heat is primarily governed by magnetic excitations, as substantiated by the experimental $C/T$ magnitude significantly exceeding the expected phonon contribution \cite{PhysRevB.105.024418} (dashed line). Furthermore, the broad maximum exhibited by $C/T$ around 1.3 K corroborates the magnetic nature of the specific heat. In the low-temperature regime, $C/T$ decreases almost linearly with $T$. A simple linear extrapolation of $C/T$ to $T\rightarrow 0$ yields a slightly negative intercept, indicating the absence or vanishingly small $T$-linear component ($\gamma T$-term) in $C(T)$. More quantitatively, we attempt to fit the $T$-dependence of $C/T$ below 0.5 K using the expression $C/T=\gamma+\alpha T+\beta T^2$ (solid line). Here the $\alpha T$-term represents magnetic excitations, and the $\beta T^2$-term accounts for magnetic excitations that deviate from the $T$-linear dependence and small but finite phonon contribution. We found that $\gamma$ is less than 5\,mJ/mol-Cu\,K$^2$, if present at all (see also Fig.\,S2). The negligibly small $\gamma$-term is corroborated by the absence of a residual term $\kappa/T$ ($T \rightarrow 0$) in thermal conductivity ($\kappa$) measurements \cite{PhysRevB.106.L220406} (inset of Fig.\,3). 

In the presence of magnetic fields, $C/T$ displays a low-temperature upturn attributed to a Schottky anomaly. This assertion is supported by the thermal conductivity measurements that probe itinerant excitations and are insensitive to the localized Schottky contribution. As shown in the inset of Fig.\,3, $\kappa/T$ measured at 10\,T coincides well with the zero-field data at low temperatures. It has been reported that the $\gamma$ value obtained by subtracting the Schottky contribution increases linearly with $H$ \cite{PhysRevB.105.L121109}. However, accurately estimating the Schottky contribution and reliably determining the $\gamma$ value in magnetic fields are challenging due to the inherent ambiguity in the subtraction process (see Fig.\,S3). The broad peak of $C/T$ at around $T=1.3$\,K is suppressed with fields and almost vanishes at 16\,T. This suppression is attributed to the formation of an excitation gap upon entering the 1/9 magnetization plateau state, which will be discussed later.

The combined analysis of the torque and specific heat data unveils the highly anomalous nature of the spin excitations in YCOB. Notably, the presence of gapless and temperature-independent spin excitations revealed by the torque measurements bears a striking resemblance to the elementary excitations in the spin channel of metals with a Fermi surface, akin to Pauli paramagnetism. It should be emphasized that this behavior is inconsistent with the anticipated characteristics ($T$-linear susceptibility) of Dirac fermions, which have been hypothesized based on the INS data and $C/T$ proportional to $T$. We examine a thermodynamic test by analyzing the data under the presumption that the elementary QPs in the QSL phase are fermionic spinons (charge-neutral fermionic QPs with $S=1/2$) possessing a Fermi surface, despite the insulating nature of the system. In this scenario, both $\chi$ and $C$ are directly related to the density of states (DOS) of the fermionic spinons. The Pauli paramagnetic susceptibility is given by $\chi_P=(1/4)(g\mu_B)^2 D(\varepsilon_F)$, where $\varepsilon_F$ is the Fermi energy, and $D(\varepsilon_F)$ is the DOS at the Fermi energy. The specific heat coefficient of the magnetic excitations is given by $\gamma_M=(1/3)(\pi k_B)^2D(\varepsilon_F)$. Provided that a substantial portion of the gapless excitations observed in the specific heat is magnetic, the $\chi_P$ and $\gamma_M$ values should result in a Wilson ratio $R_W $ close to unity, a hallmark of a Fermi surface. Remarkably, assuming conventional fermions and $\chi_P = 6.5\times10^{-3}$\,emu/mol-Cu determined by the magnetic torque experiments, the estimated $\gamma_M$ value is approximately 450\,mJ/K$^2$mol$^{-1}$, at least 90 times larger than the maximum $\gamma_M=$5\,mJ/K$^2$mol$^{-1}$ obtained from the specific heat measurements. Although ferromagnetic fluctuations may account for this discrepancy, such ferromagnetic fluctuations are absent in YCOB (see Supplemental Material). The presence of local ferromagnetic clusters has been proposed based on the $H$-linaer dependence of $M/\mu_0H$ \cite{shivaram2024non}. However, such a $H$-linear behavior of $M/\mu_0H$ is not observed in our sample (see Fig.\,S4). It is worth noting that in a QSL state of the organic triangular AFM EtMe$_3$Sb[Pd(dmit)$_2$]$_2$, Pauli-paramagnetic-like $\chi$ and finite $\gamma$ with a Wilson ratio close to unity have been reported \cite{yamashita2011gapless,watanabe2012novel}. This metallic-like behavior in the insulating material has been discussed in the context of fermionic spinons forming a Fermi surface. The present results demonstrate that the nature of the 2D QSL state in the kagome lattice is essentially different from that in the triangular lattice.

Given the inadequacy in elucidating the observed phenomena within the framework of fermionic excitations, we propose that the low-energy excitations of YCOB are described by bosonic QPs \cite{PhysRevLett.66.1773,PhysRevB.45.12377,PhysRevB.74.174423,punk2014topological,PhysRevB.50.258,PhysRevB.37.4936}. In this scenario, while $C$ is directly contingent upon the DOS of the QPs, $\chi$ is directly related to the number of bosonic QPs. The finite $\chi$ even at $T\rightarrow 0$ indicates that the compressibility of bosonic QPs remains finite. Since $M$ is proportional to the number of bosonic QPs, the $H$-linear increase of $M$ even at very low fields and $T \rightarrow 0$ indicates that the number of the bosonic QPs is finite, pointing to the emergence of the superfluid state of bosonic QPs \cite{PhysRevB.50.258,PhysRevB.37.4936}.

We proceed to discuss the phase diagram in the framework of the bosonic spinons proposed in Ref.\,\citenum{PhysRevB.50.258} [Fig.\,4(a)]. At very low fields, a paramagnetic state crosses over to a gapped QSL state (dashed line) with decreasing $T$. The tiny spin gap, sufficiently smaller than 0.16\,K ($\sim J/500$), is closed by very small magnetic fields $\mu_0 H_c\ll 100$\,mT, leading to the transition of bosonic spinons into a superfluid state (solid line). A characteristic feature of the superfluid state is the presence of a gapless momentum-linear Goldstone mode [Fig.\,4(b)], which gives rise to the observed $T$-linear dependence of $C/T$. It is worth noting that a gapless Goldstone mode associated with Bose-Einstein condensation of triplet excitations has been reported in the spin dimer compound TlCuCl$_3$ \cite{ruegg2003bose}. Because of the 2D nature of the kagome AFM, this superfluid transition is a Kosterlitz-Thouless transition. Thus, in the presence of very small fields, the ground state is a gapless state with algebraic spin-spin correlations. In this scenario, the cone-like continuum excitations in YCOB observed by INS in zero field \cite{zeng2023dirac} are attributed to the bosonic spinon excitations. To confirm this scenario, it is necessary to detect the extremely tiny spin gap, which is undetectable in the present $T$- and $H$- ranges, warranting future study.

Intense magnetic fields induce a profound transformation in the quantum states of YCOB, manifested by plateau-like magnetization anomalies at fractional values of 1/9 and 1/3 of the fully polarized state \cite{suetsugu2023emergent,jeon2024,zheng2023unconventional} [red line in Fig.\,4(c)]. However, the nature of the 1/9 magnetization anomaly remains a subject of ongoing debate. While some groups assert that it is a genuine plateau with a distinct energy gap \cite{suetsugu2023emergent}, others propose that it arises from quantum oscillations originating from gapless Dirac spinons \cite{zheng2023unconventional}.

To address this issue, we performed measurements of the magnetocaloric effect (MCE) \cite{kihara2013adiabatic}. The MCE can be quantified by the isothermal entropy change associated with the application or removal of a magnetic field. These quantities are related to the magnetic entropy change, and hence allow for the verification of the emergence of an energy gap. Figure\,4(c) depicts the contour plot of the magnetic entropy $S_M$ extracted from the isentropic $T(H)$ data (black solid lines) in pulsed magnetic fields up to 60\,T, obtained from MCE measurements conducted under quasi-adiabatic conditions. The $T(H)$ curves measured during the increasing and decreasing magnetic field sweeps exhibit good agreement and reflect a constant entropy contour. In the low-temperature and high-field regime, approaching the 1/3 magnetization plateau, $T(H)$ increases and exhibits a maximum, implying that $S_M$ becomes minimal. As the formation of an energy gap reduces $S_M$, this result is consistent with the emergence of the 1/3 magnetization plateau. At lower fields, $T(H)$ exhibits a distinct peak with a maximum (corresponding to a minimum in $S_M$) at approximately 20\,T, indicating the formation of an energy gap. Notably, this field is very close to the field associated with the 1/9 magnetization anomaly \cite{suetsugu2023emergent}. Thus, the MCE results unequivocally corroborate the manifestation of an energy gap in the vicinity of 20\,T, providing compelling thermodynamic evidence for the emergence of the 1/9 fractional magnetization plateau phase.
 
The present experimental findings unveil that the ground state of the kagome AFM YCOB in the absence or under extremely weak fields is characterized by gapless magnetic excitations of a likely bosonic nature. This is in stark contrast to the QSL state of the triangular lattice with fermionic spinons. Furthermore, the gapless ground state undergoes a quantum phase transition to a gapped plateau state upon the application of intense fields. These observations elucidate the long-standing enigmas surrounding the ground state of the kagome AFM, providing crucial insights into the highly quantum entangled state manifested in quantum magnets.

\begin{acknowledgments} 
We thank C. Hotta, K. Penc, M. G. Yamada, M. Tokunaga, N. Kawakami, and S. Fujimoto for insightful discussions. We acknowledge K. Mehlawat and Y. Zhuo for their help during the MCE experiments. This work is supported by Grants-in-Aid for Scientific Research (KAKENHI) (Nos. 23K13060 and 23H00089) and Transformative Research Areas A “Extreme Universe” (No. 24H00965) from the Japan Society for the Promotion of Science, and JST CREST (JPMJCR19T5). The work in Wuhan was supported by the National Natural Science Foundation of China (No. 12274153).
\end{acknowledgments} 

\clearpage
\begin{figure}
	\includegraphics[clip,width=8.5cm]{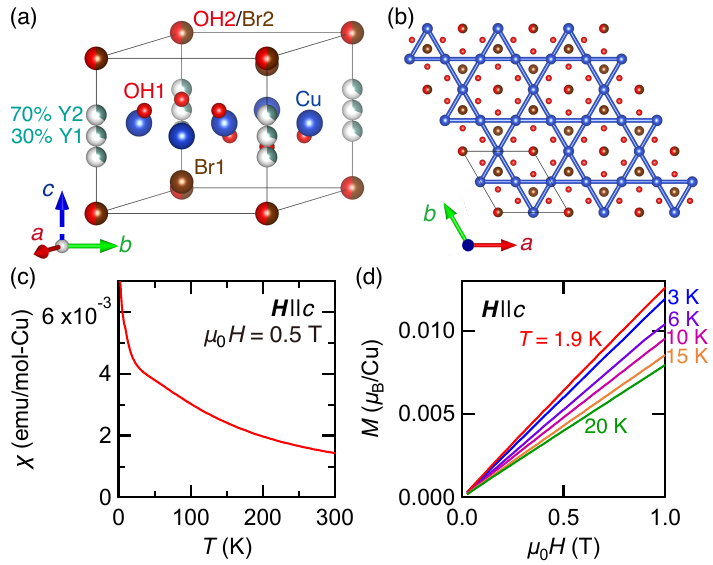}
	\caption{Crystal structure and magnetic properties of YCOB. (a) Crystal structure of YCOB. The intersite mixing of OH2 and Br2 displaces 70\% of Y$^{3+}$ ions from the optimal Y1 position on the kagome plane. (b) The kagome plane of Cu$^{2+}$ ions in YCOB. YCOB features of a 2D perfect kagome lattice of Cu$^{2+}$ ions without intersite mixing between magnetic Cu$^{2+}$ and non-magnetic Y$^{3+}$ ions due to the significant mismatch in their ionic radii. (c) Temperature dependence of magnetic susceptibility $\chi(T)$ for $\bm{H}||c$ at 0.5\,T. A steep increase of $\chi(T)$ below $\sim30$\,K is attributed to, in part, a Curie-like contribution from impurity spins. (d) Field dependence of magnetization $M$ for $\bm{H}||c$. The magnetization curve $M(H)$ shows a strong temperature dependence down to 1.9\,K.
	\label{fig:crystal}
	}
\end{figure}

\begin{figure*}
	\includegraphics[clip,width=16cm]{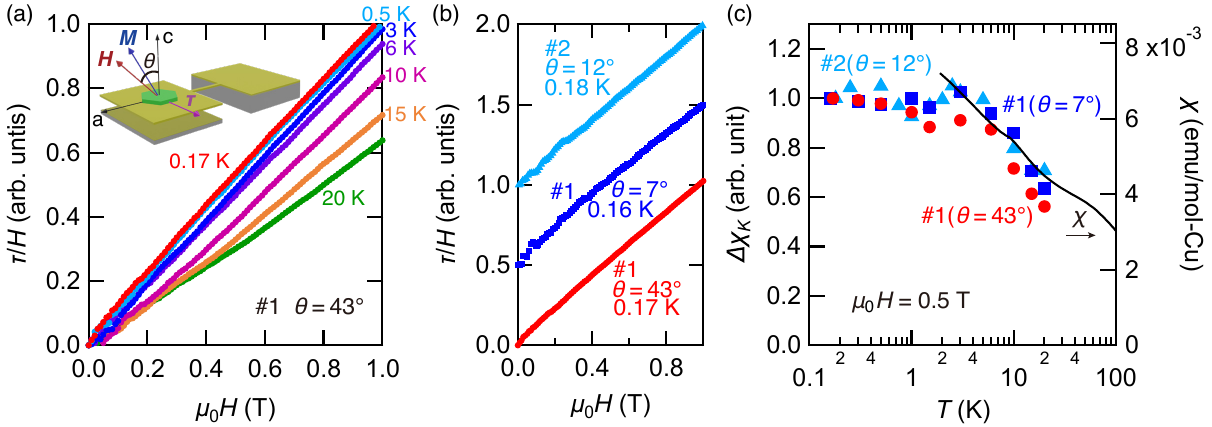}
	\caption{Magnetic torque of YCOB. (a) Field dependence of magnetic torque divided by magnetic field, $\tau/H$, for sample \#1 at $\theta = 43^\circ$. The slope of $\tau/H$ remains nearly $T$-independent below 6\,K, indicating the Curie-like contribution from impurities is almost completely canceled out. Even at $T=$170\,mK, $\tau/H$ shows $H$-linear increase down to at least 50\,mT, demonstrating the spin gap, if it exists, is well below 170\,mK $\sim J/500$. The inset shows the experimental setup for magnetic torque measurements using a capacitive cantilever. (b) Field dependence of $\tau/H$ for sample \#1 at $\theta = 43^\circ$ and $T=0.17$\,K (red circles), sample \#1 at $\theta = 7^\circ$ and $T=0.16$\,K (blue squares), and sample \#2 at $\theta = 12^\circ$ and $T=0.18$\,K (light blue triangles). The latter two datasets are vertically shifted for clarity. For all data, the magnitude of $\tau/H$ is normalized by the value at $\mu_0 H=1$\,T. (c) Temperature dependence of $\Delta\chi_K (\propto \tau/H^2)$ at $\mu_0 H=$0.5\,T. The value of $\Delta\chi_K$ is normalized by the value at the lowest measured temperature. For all data, $\Delta\chi_K$ increases with decreasing $T$ and then saturates below 5\,K, demonstrating that the intrinsic susceptibility from the kagome plane $\chi_K$ remains finite down to very low temperature. The finite value of $\chi_K$ is estimated to be $6.5\times10^{-3}$\,emu/mol-Cu by comparing $\Delta\chi_K$ with $\chi$ (solid line) measured by SQUID magnetometry at $\mu_0H = 0.5$\,T. The data for $\Delta\chi_K$ and $\chi$ above 5\,K are adjusted such that the value of $\Delta\chi_K$ is well scaled with that of $\chi$. At lower temperatures, $\chi$ continues to increase with decreasing $T$ and deviates from $\Delta\chi_K$ due to the presence of the Curie-like contribution in $\chi$.
	\label{fig:torque}
	}
\end{figure*}

\begin{figure}
	\includegraphics[clip,width=7cm]{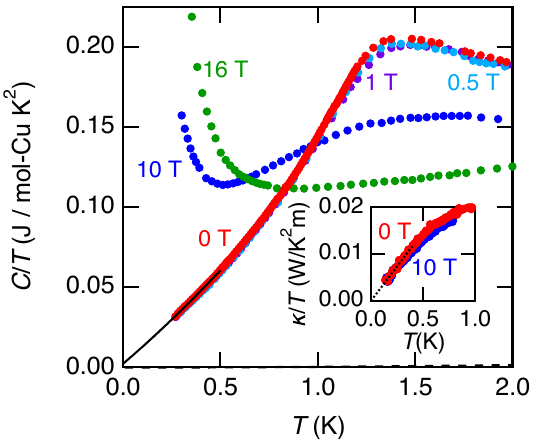}
	\caption{Temperature dependence of specific heat divided by temperature $C/T$ of YCOB. The magnitude of $C/T$ is much larger than that expected for the phonon contribution \cite{PhysRevB.105.024418} (dashed line). The fit of zero-field data below 0.5\,K using $C/T = \gamma + \alpha T + \beta T^2$ (solid line) yields a small value of $\gamma=2.3$\,mJ/mol-Cu K$^2$. This value is substantially smaller than 450\,mJ/mol-Cu K$^2$ estimated from $\chi_K = 6.5\times 10^{-3}$\,emu/mol-Cu, assuming the Wilson ration $R_W$ is close to unity. In magnetic fields, the Schottky anomaly gives rise to an upturn in $C/T$ at low temperatures. The inset depicts the temperature dependence of thermal conductivity divided by temperature $\kappa/T$. The dotted line in the inset indicates a linear fit of $\kappa/T$ at low temperatures, which gives a negligibly small residual term $\kappa/T$ ($T\rightarrow 0$).
	\label{fig:specifc_heat}
	}
\end{figure}

\begin{figure}
	\includegraphics[clip,width=8cm]{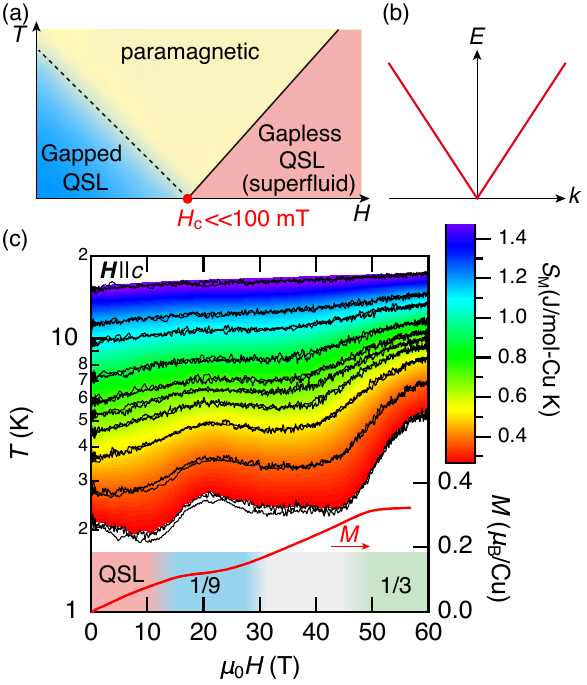}
	\caption{Phase diagram and MCE experiments of YCOB. (a) Possible phase diagram based on the bosonic spinon scenario discussed in Ref.\,\citenum{PhysRevB.50.258}. The dashed line represents a crossover from the paramagnetic state to the gapped QSL state with decreasing $T$. The tiny spin gap is closed by a small magnetic field $H_c \ll 100$\,mT, leading to a Kosterlitz-Thouless transition (solid line) to the superfluid state of bosonic spinons. (b) Goldstone mode in the superfluid state. The energy $E$ of the gapless Goldstone mode increases linearly with momentum $k$ (red lines). (c) MCE measurements performed under quasi-adiabatic conditions up to 60\,T. The contour plot of the magnetic entropy $S_M$ is extracted from the isentropic $T(H)$ data (black solid lines) obtained from MCE experiments and the temperature dependence of $S_M$ in zero field (see Fig.\,S5). The red solid line indicates the magnetization curve at $T=0.6$\,K reported in Ref.\,\citenum{suetsugu2023emergent}. At low temperatures, the $T(H)$ curves exhibit maxima around 20\,T and 60\,T, demonstrating the formation of energy gaps in the 1/9 and 1/3 plateau regions.
	\label{fig:MCE}
	}
\end{figure}

\clearpage
\bibliography{ref.bib}

\end{document}



\title{Supplemental Material for Gapless spin excitations in a quantum spin liquid state of $\bm{S}$=1/2 perfect kagome antiferromagnet}


\author{S. Suetsugu}
\affiliation{Department of Physics, Kyoto University, Kyoto 606-8502, Japan}
\author{T. Asaba}
\affiliation{Department of Physics, Kyoto University, Kyoto 606-8502, Japan}
\author{S. Ikemori}
\affiliation{Department of Physics, Kyoto University, Kyoto 606-8502, Japan}
\author{Y. Sekino}
\affiliation{Department of Physics, Kyoto University, Kyoto 606-8502, Japan}
\author{Y. Kasahara}
\affiliation{Department of Physics, Kyoto University, Kyoto 606-8502, Japan}
\author{K. Totsuka}
\affiliation{Yukawa Institute for Theoretical Physics, Kyoto University, Kyoto 606-8502, Japan}
\author{B. Li}
\affiliation{Wuhan National High Magnetic Field Center and School of Physics, Huazhong University of Science and Technology, 430074 Wuhan, China}
\author{Y. Zhao}
\affiliation{Wuhan National High Magnetic Field Center and School of Physics, Huazhong University of Science and Technology, 430074 Wuhan, China}
\author{Y. Li}
\affiliation{Wuhan National High Magnetic Field Center and School of Physics, Huazhong University of Science and Technology, 430074 Wuhan, China}
\author{Y. Kohama}
\affiliation{Institute for Solid State Physics, The University of Tokyo, Kashiwa, Chiba 277-8581, Japan}
\author{Y. Matsuda}
\affiliation{Department of Physics, Kyoto University, Kyoto 606-8502, Japan}
\date{\today}


\maketitle
%
\section{Method}
\subsection{Single crystal growth and characterization}
Single crystals of YCOB were grown by a hydrothermal method as reported previously \cite{PhysRevB.105.024418}. Magnetization experiments up to 7\,T were conducted using a commercial SQUID magnetometer. For magnetization measurements, multiple single crystals aligned along the crystal $c$ axis were collected.

\subsection{Magnetic torque experiments}
Magnetic torque experiments were performed by a capacitive cantilever method \cite{li2011coexistence,PhysRevB.90.064417}. We measured two single crystals, sample \#1 and \#2, by applying fields $\bm{H}$ at an angle $\theta$ relative to the $c$-axis. Each sample was mounted on one end of a beryllium copper cantilever with the crystal $c$ axis facing up.  We deposited a gold film on a sapphire plate and then placed it under the sample end of the cantilever, as shown in the inset of Fig.\,2(a). The other end of the cantilever was glued to another sapphire plate. The magnetic torque was tracked by measuring the capacitance between the cantilever and the gold film.

\subsection{Specific heat experiments}
Specific heat experiments were conducted by a long-relaxation time method \cite{PhysRevB.63.094508,PhysRevLett.99.057001} by applying field $H$ along the crystal $c$ axis. The magnetic contribution $C_M/T$ is obtained by subtracting the phonon contribution (dashed line) reported in Ref.\,\citenum{PhysRevB.105.024418}. Then, the magnetic entropy $S_M$ was obtained by integrating $C_M/T$ (see Fig.\,S5).

\subsection{Thermal conductivity experiments}
Thermal conductivity $\kappa$ was measured by a standard steady-state method by applying a temperature gradient $\bm{q}$ along a certain direction within the $ab$ plane. Three gold wires were attached using silver paste to serve as heat links to a 10-$\mathrm{k\Omega}$ chip resistor as a heater and two field-calibrated thermometers. One end of the crystal was glued to a silver plate acting as a heat bath using silver paste. The magnetic field $\bm{H}$ was applied along the crystal $c$ axis.

\subsection{Magnetocaloric effect experiments}
MCE experiments were performed up to 60\,T under quasi-adiabatic conditions \cite{kihara2013adiabatic}. An Au$_{16}$Ge$_{84}$ thin film was sputtered on the surface of a single crystal to serve as a thermometer. For the MCE experiments, magnetic fields were generated using a non-destructive pulse magnet installed at the International MegaGauss Science Laboratory of the Institute for Solid State Physics at the University of Tokyo. 

\section{Discrepancy in fermionic spinon scenario}
Here we discuss why the observed behavior of magnetic torque and specific heat cannot be understood by the fermionic spinon scenario, which has been widely argued in kagome systems. Within the framework of Fermi liquid theory, the magnetic susceptibility $\chi$ is renormalized by the Landau parameter $F_0$, which encapsulates the effects of quasiparticle (QP) interactions arising from the exchange of spin fluctuations. In the presence of ferromagnetic fluctuations, $F_0$ acquires negative values, which amplify the QP magnetic response, resulting in an augmented susceptibility, expressed as $\chi \propto 1/(1+F_0)$. This enhancement of $\chi$ leads to a divergent behavior of the Wilson ratio $R_W$ in the vicinity of the ferromagnetic quantum critical point. While $\chi$ diverges, the Sommerfeld coefficient $\gamma$ remains finite. This disparity between $\chi$ and $\gamma$ manifests as a divergence in $R_W$. However, the presence of such ferromagnetic fluctuations in YCOB can be excluded for the following reasons. In the vicinity of the ferromagnetic order, an external field $H$ induces the saturation of the DC magnetization at low temperatures. In such a case, the magnetic torque $\tau$ should vanish or exhibit saturating behavior at low fields. However, $\tau/H$ continues to increase with $H$ up to 13\,T even at the lowest measured temperature [see Fig.\,S1(c)].

\begin{figure*}
	\includegraphics[clip,width=16cm]{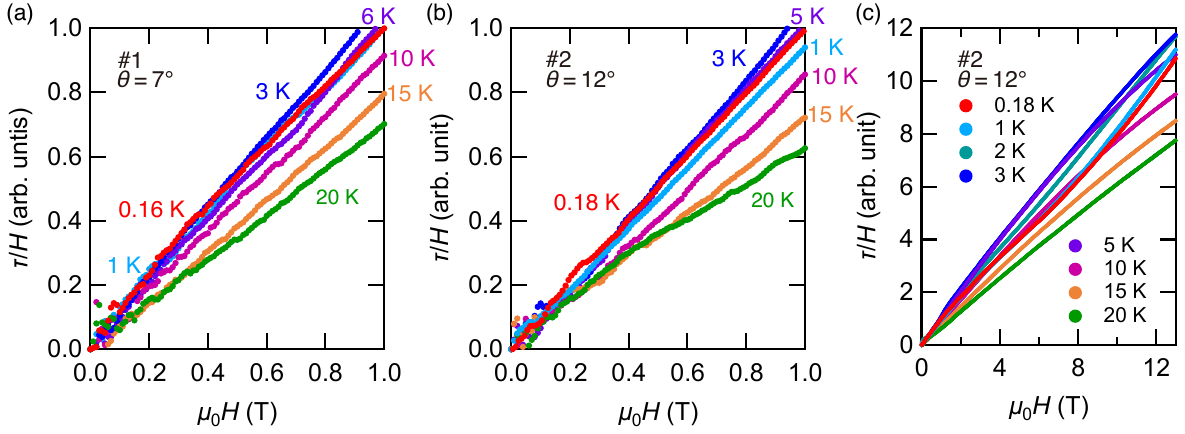}
	\caption{Magnetic torque of YCOB. (a) and (b) Field dependence of magnetic torque divided by temperature $\tau/H$ for sample \#1 at $\theta = 7^\circ$ (a) and sample \#2 at $\theta = 12^\circ$. For both samples, $\tau/H$ increases linearly with $H$, and its slope remains nearly $T$-independent up to 6\,K. (c) Field dependence of $\tau/H$ at higher fields for sample \#2 at $\theta = 12^\circ$. For all measured temperatures, $\tau/H$ continues to increase with $H$ up to 13\,T.
	}
	\label{fig:torque_add}
\end{figure*}

\begin{figure*}
	\includegraphics[clip,width=12cm]{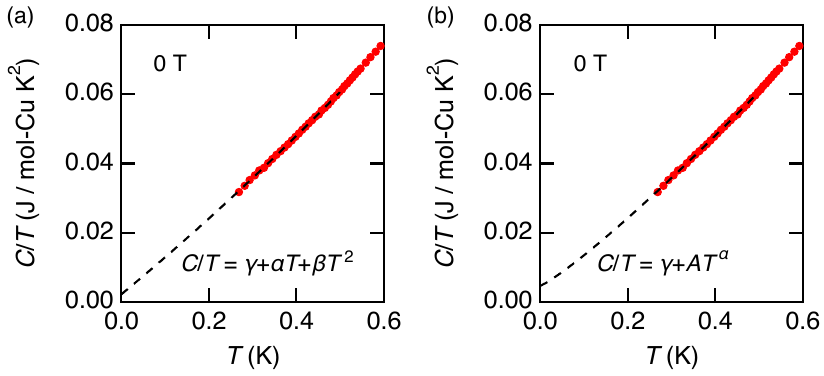}
	\caption{Temperature dependence of specific heat divided by temperature $C/T$ in zero field. The dashed lines represent fits to the data below $T<0.5$\,K using $C/T = \gamma + \alpha T + \beta T^2$ (a) and $C/T = \gamma + AT^\alpha$ (b), yielding small values of $\gamma =$ 2.3 and 4.7\,mJ/mol-Cu K$^2$, respectively. 
	}
	\label{fig:C_0T}
\end{figure*}

\begin{figure*}
	\includegraphics[clip,width=12cm]{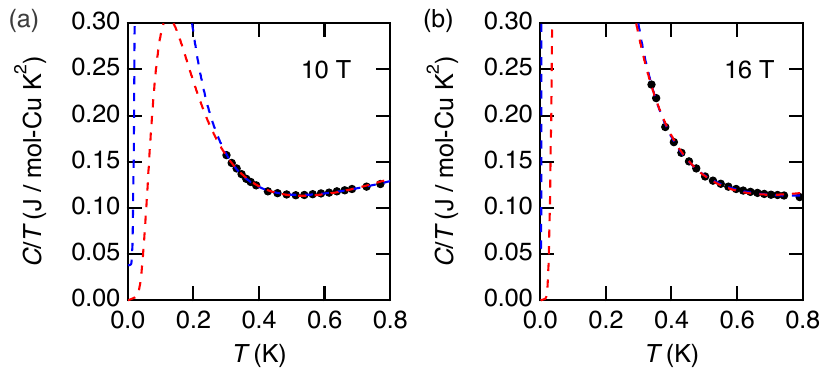}
	\caption{Temperature dependence of specific heat divided by temperature $C/T$ in magnetic fields. (a) and (b) Temperature dependence of $C/T$ at 10\,T (a) and 16\,T (b). The blue dashed lines represent fits to $C/T = \gamma + \alpha T + \beta T^2 + C_\mathrm{Sch}/T$ including the $\gamma$-term, while the red dashed lines indicate fits using $C/T = \alpha T + \beta T^2 + C_\mathrm{Sch}/T$. Here $C_\mathrm{Sch} = A_\mathrm{Sch}(\frac{\Delta}{k_BT})^2 \exp(-\frac{\Delta}{k_BT})$ is the two-level Schottky specific heat, where $A_\mathrm{Sch}$ is a constant determined by the number of two-level systems and $\Delta$ is the energy gap between the two levels. Since both formulas, with and without the $\gamma$ term, fit the data very well, determining a reliable $\gamma$ value in magnetic fields is challenging.
	}
	\label{fig:Schottky}
\end{figure*}

\begin{figure*}[h]
	\includegraphics[clip,width=7cm]{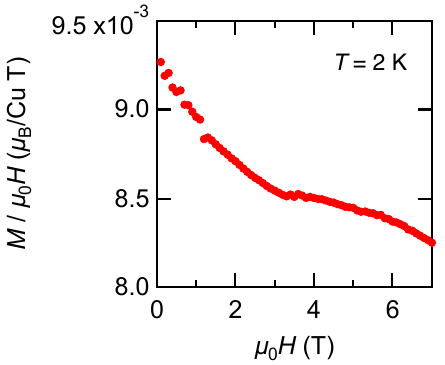}
	\caption{Field dependence of magnetization divided by $\mu_0 H$ of YCOB. The $H$-linear behavior reported in Ref.\,\citenum{shivaram2024non} is not observed in our sample. 
	}
	\label{fig:M_B}
\end{figure*}

\begin{figure*}
	\includegraphics[clip,width=12cm]{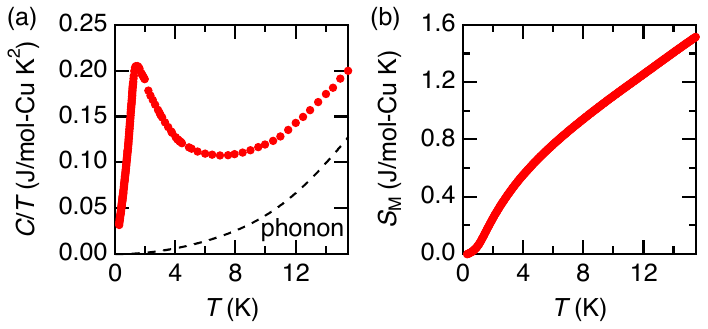}
	\caption{Temperature dependence of magnetic entropy of YCOB. (a) Temperature dependence of $C/T$ in zero field. The magnetic contribution $C_M/T$ is obtained by subtracting the phonon contribution (dashed line) reported in Ref.\,\citenum{PhysRevB.105.024418}. (b) Temperature dependence of magnetic entropy $S_M$ in zero field.
	}
	\label{fig:entropy}
\end{figure*}

\clearpage

\bibliography{ref.bib}